\documentstyle[preprint,prb,aps]{revtex}
\newcommand{\ave}[1]{\mbox{$\langle #1 \rangle$}}

\newcommand{\beq}{\begin{equation}}
\newcommand{\eeq}{\end{equation}}
\newcommand{\beqa}{\begin{eqnarray}}
\newcommand{\eeqa}{\end{eqnarray}}
\newcommand{\bmath}{\begin{mathletters}}
\newcommand{\emath}{\end{mathletters}}

\begin{document}
\draft


\title{Quantum versus Semiclassical Description of Selftrapping: 
Anharmonic Effects}
\author{S. Raghavan$^1$,
 A. R. Bishop$^2$, and V. M. Kenkre$^3$}
\address{
$^1$ International Centre for Theoretical Physics, 34100 Trieste, Italy and\\
Rochester Theory Center for Optical Science and Engineering,\\ Department of
Physics \& Astronomy, University of Rochester, Rochester NY
14627\cite{byline1}, \\
$^2$ Los Alamos National Laboratory, Los Alamos NM 87545\\
$^3$ Center for Advanced Studies and Dept. of Physics \& Astronomy \\
University of New Mexico, Albuquerque NM 87131}
\maketitle


\begin{abstract}
Selftrapping has been traditionally studied on the assumption that
quasiparticles interact with harmonic phonons and that this interaction is
linear in the displacement of the phonon. To complement recent 
semiclassical
studies of anharmonicity and nonlinearity in this context, we present below
a fully quantum mechanical analysis of a two-site system, where the
oscillator is described by a tunably anharmonic potential, with a square
well with infinite walls and the harmonic potential as its extreme limits,
and wherein the interaction is nonlinear in the oscillator displacement. We
find that even highly anharmonic polarons behave similar to their harmonic
counterparts in that selftrapping is preserved for long times in the limit
of strong coupling, and that the polaronic tunneling time  scale depends
exponentially on the polaron binding energy. Further, in agreement, with
earlier results related to harmonic polarons, the semiclassical
approximation agrees with the full quantum result in the massive oscillator
limit of small oscillator frequency and strong quasiparticle-oscillator
coupling.
\end{abstract}
\section{Introduction}

\label{sec:intro} Recent work by Grigolini and collaborators \cite{Bonci2},
and by Salkola and the present authors \cite{sbkr,krbs-rkbs}, have 
uncovered
subtle features associated with selftrapping of quasiparticles in
interaction with vibrations. The vibrations considered in all those 
analyses
have been harmonic. The question of how polaron dynamics and selftrapping
are affected by anharmonicities in the vibrations was raised by Kenkre 
\cite
{samos} several years ago at the level of the discrete nonlinear
Schr\"{o}dinger equation (DNLSE) and analyzed by Kenkre and collaborators 
in
the context of rotational polarons\cite{samos,kwh-wk}, exponential
saturation \cite{occassnot,cret-rotpla-ak}, and general considerations 
\cite
{cret-rotpla-ak}. Since the validity of the DNLSE has been called into
question by recent considerations \cite{Bonci2,sbkr,krbs-rkbs}, it is
important to examine the issue of what polarons, or selftrapping, owe to
harmonic features from a starting point which is fully quantum. The present
paper is devoted to such an examination for a two-site system.

We will focus here on confined systems rather than on periodic systems such
as those which may lead to rotational polarons \cite{samos,cret-rotpla-ak}.
In a certain sense, the most anharmonic potential conceivable is that which
corresponds to a box with infinitely high walls as it corresponds to a
harmonic piece with vanishing frequency throughout the interior of the box
but one with infinite frequency at the wall. We choose the symmetric %
\mbox{P\"oschl-Teller  } potential given by 
\begin{equation}
V_{PT}(x)=U_{0}\tan ^{2}(ax),  \label{eq:ptpotl}
\end{equation}
because it allows continuous transition between the harmonic oscillator and
the box limits and because it can be treated analytically with ease. In 
Eq.~(%
\ref{eq:ptpotl}) $U_{0},a$ are constants defining, respectively, the
strength and confining region of the potential. The potential becomes
infinitely steep at $x=\pm \pi /2a$. By rewriting the strength of the
potential $U_{0}$ as $\lambda (\lambda -1)\hbar ^{2}a^{2}/2m$ where $m$ is
the mass of the particle, we introduce the parameter $\lambda $ which
describes the departure of the potential between the box and harmonic
oscillator limits. In the limit $\lambda \rightarrow 1$, the %
\mbox{P\"oschl-Teller  } potential becomes the infinite square well of 
width 
$\pi /a$. In the opposite limit $\lambda \rightarrow \infty ,a\rightarrow
0,\lambda a^{2}$ remaining constant (and finite), one recovers the harmonic
oscillator potential. The eigenenergies of the \mbox{P\"oschl-Teller  }
potential (\ref{eq:ptpotl}) are given by \cite{nieto2} 
\begin{equation}
E_{n}=\frac{\hbar ^{2}a^{2}}{2m}(n^{2}+2n\lambda +\lambda
),\;~\;~\;~n=0,1,2,...  \label{eq:ptpotlerg}
\end{equation}
and the corresponding eigenfunctions are 
\begin{equation}
\phi _{n}(x)\equiv \mbox{$\langle x|\phi_n \rangle$}=N_{n}\cos
^{1/2}axP_{n+\lambda -1/2}^{1/2-\lambda }(\sin ax),  \label{eq:}
\end{equation}
where $P_{\alpha }^{\beta }(t)$ are the associated Legendre functions, 
with $%
N_{n}=\left( \frac{a(n+\lambda )\Gamma (n+2\lambda )}{\Gamma (n+1)}\right)
^{1/2}$.

\section{Anharmonic Polaron - Stationary aspects}

\label{sec:sec2} Consider a two-site system consisting of a quasiparticle,
like an electron or an exciton, whose inter-site hopping is described by a
matrix element of strength V. The quasiparticle also strongly interacts 
with
a vibrational mode between the two sites. This vibrational mode is 
described
by the \mbox{P\"oschl-Teller  } potential (\ref{eq:ptpotl}). In the 
harmonic
case the usual interaction is linear in the vibrational amplitude and
consequently connects nearest neighbour eigenstates of the oscillator. 
These
two features are distinct from each other. In developing a scheme for
analyzing effects of a genealization to anharmonic situations,  we must
maintain either one or the other of the two features. We have studied both
cases. Here we present  results of maintaining the second feature, viz., an
interaction which joins nearest neighbour energy eigenstates. As Nieto and
Simmons \cite{nieto2} and Crawford and Vrcsay \cite{craw-vrc} have pointed
out in a different context, a sinusoidal interaction posseses this feature
for the \mbox{P\"oschl-Teller  } potential, and also reduces to the linear
form in the harmonic limit. We thus take the full Hamiltonian of our system
to be 
\begin{equation}\label{eq:ham}
H=\frac{\omega _{0}}{2(\lambda +1/2)}\left( \hat{\pi _{z}}^{2}+\lambda
(\lambda -1)\tan ^{2}\hat{z}\right) +\omega _{0}g\sqrt{\left( \lambda 
+\frac{%
1}{2}\right) }\hat{p}\,\sin \hat{z}+V\hat{r},
\end{equation}
where 
\begin{equation}
\omega _{0}=\frac{a^{2}}{m}(\lambda +\frac{1}{2})  \label{eq:defnomega}
\end{equation}
is the difference between the energies of the first excited state and the
ground state of the \mbox{P\"oschl-Teller  } potential, $z$ is the
dimensionless oscillator coordinate $ax$, and $g$ is the
quasiparticle-oscillator coupling constant. Here and henceforth we put $
\hbar =1$ for simplicity.

The operators $\hat{p},\hat{r}$ are the operators describing the
quasiparticle, with $\hat{p}=c_{1}^{\dagger }c_{1}-c_{2}^{\dagger 
}c_{2},\;%
\hat{r}=(c_{1}^{\dagger }c_{2}+c_{2}^{\dagger }c_{1})$, where the $c$'s are
quasiparticle creation and destruction operators. The factorization or the
semiclassical approximation (SCA) consists of assuming, equivalently, that
the oscillator operators behave classically or that products of
quasiparticle-oscillator operators can be factorized.

We compare the SCA with the fully quantum mechanical results first by
computing the polaron binding energies. This is done easily in the
strong-coupling limit by freezing the quasiparticle hopping dynamics. We
note first that, in the harmonic oscillator limit, the polaron binding
energy is proportional to $g^{2}$ whereas, in the opposite limit of the
infinite square well, the width of the well remains finite and the
interaction produces a lowering of energy that is proportional to $g$. This
cross-over behaviour becomes evident from the full quantum-mechanical
calculations as shown in Fig.~1. Plotted in the main figure is the binding
energy (in arbitrary units) as a function of $g$. In the inset, the same
quantities are plotted on a logarithmic scale. The bold lines indicate the
fully quantum-mechanical calculation and the dashed lines indicate the
results of the SCA. In all the cases, the oscillator frequency $\omega 
_{0}$
has been kept fixed, and $\lambda $ varied allowing the oscillator to pass
smoothly from the box ($\lambda =1$) limit to the harmonic ($\lambda
\rightarrow \infty $) limit. Two key results are evident in Fig.~1. First, 
the SCA results agree with the exact ones only in the harmonic oscillator
limit; in the square-well limit, the departure becomes quite drastic.
Second, (see inset), when the system is in the box-limit ($\lambda =1$), 
the
fully quantum system behaves harmonically for small $g$ but exhibits a
cross-over for larger values of $g$ showing the true box-limit slope of
unity.

We also calculate the overlap between the adiabatically displaced
ground-state wavefunctions. This overlap, basically the Huang-Rhys factor, 
governs the polaronic tunneling rate in the strong-coupling limit. One
knows, for instance, that in the harmonic oscillator limit, the tunneling 
rate is proportional to $e^{-g^{2}}$. In Fig.~2, we plot the overlap factor
(logarithmically) as a function of $g$ for  the box limit, i.e., $\lambda 
=1$
(dashed line), and for the harmonic oscillator limit, i.e., $\lambda
\rightarrow \infty $ (solid line). The quadratic dependence is clearly seen
for the latter. However, for $\lambda =1$, the rise is sublinear, showing
that the dependence of the overlap factor on $g$ is much weaker.

\section{Heisenberg equations}

We discuss in this section the temporal evolution of the system, and show
how the SCA differs from the fully quantum mechanical treatment. The
equations of motion corresponding to the Hamiltonian (\ref{eq:ham}) can be
written as
\bmath
\label{eq:heisen}
\begin{eqnarray}
\dot{\hat{p}} &=&2V\hat{q}  \\
\dot{\hat{q}}
 &=&-2V\hat{p}+2\omega _{0}g(\lambda +1/2)^{1/2}\hat{r}\sin \hat{z} \\
\dot{\hat{r}}
 &=&-2\omega _{0}g(\lambda +1/2)^{1/2}\hat{q}\sin \hat{z}  
\\
\dot{\hat{z}}
 &=&\frac{\omega _{0}}{(\lambda +1/2)}\hat{\pi }_{z} \\
\dot{\hat{\pi }_{z}} &=&-\omega _{0}\left( \frac{\lambda (\lambda -1)}{%
(\lambda +1/2)}\tan \hat{z}\sec ^{2}\hat{z}+g(\lambda 
+1/2)^{1/2}\hat{p}\cos 
\hat{z}\right) ,
\end{eqnarray}
\emath
where $\hat{q}=i(c_1^\dagger c_2-c_2^\dagger c_1)$ and 
the quasiparticle operators $\hat{p},\hat{q},\hat{r}$ cyclically
satisfy the commutation relations, $[\hat{p},\hat{q}]=2i\hat{r}$, and 
$[\hat{%
z},\hat{\pi _{z}}]=i$. As stated earlier, the SCA consists in assuming the
oscillator operators $\hat{z},\hat{\pi _{z}}$ to be $c$-numbers.  In the
temporal analysis, we compare the results of such an approximation with
those given by the full quantum evolution described by 
Eqs.~(\ref{eq:heisen}).
 We plot in Figs.~3-5, the evolution of the population difference of the
quasiparticle between the two sites $p(t)$ as a function of dimensionless
time $Vt$. In all our calculations, the initial condition used for the
quantum system is the ground state of the quasiparticle-oscillator system
projected onto the one-site localized part of the Hilbert space, such that
$\ave{\hat{p}}(0)=1.$ The initial condition used for the SCA calculation is
$p(0)=1,\dot{z}(0)=\dot{\pi}_z(0)=0$. 
In all the plots, the solid line indicates the full quantum
evolution and the dashed line indicates the evolution due to the SCA. The
polaron binding energy has been kept constant to facilitate comparison. 
This
value (which we take to be 1.5 V) equals $g^{2}\omega _{0}/2$ in the
harmonic oscillator limit. In Fig.~3, $\lambda =200$, the oscillator
potential is essentially that of the harmonic oscillator potential. The
oscillator energy $\omega _{0}$ takes on the values $10V,V,0.1V$ in 
(a),(b),
and (c) respectively. As discussed elsewhere \cite{sbkr}, whereas the SCA
shows self-trapping for all the values of the oscillator frequency, the 
full
quantum evolution differs substantially, except in the limit of low
oscillator energy $\omega _{0}=0.1V$. In this limit, for short times, the
full quantum evolution and SCA agree in that both show self-trapping, with
nearly the same average value of self-trapping and oscillation frequency.
However, the quantum evolution shows a considerably richer structure
involving collapses and revivals. At much longer times, the dressed
quasiparticle tunnels from one site to the other. This is evident in Fig.~3
(b). When $\lambda =10$, (Fig.~4), the potential is more `square-well'-like
and some departures, especially in the small oscillator frequency regime
(Fig.~4 (b)) are visible. For instance, the `silent runs' separating the
collapse and revival sequence, are less quiescent, and the agreement 
between
the SCA and the full quantum evolution is slightly worse. In Fig.~5, we 
take 
$\lambda =1$, wherein the potential is essentially the infinite 
square-well.
Whereas the agreement between the SCA and the full quantum evolution is 
best
when the oscillator energy is least, $\omega _{0}=0.1V$, (Fig.~5 (c)), the
agreement is far worse than for the harmonic potential (Fig.~5 (c)).
Further, the `silent runs' are barely noticeable, with the collapses and
revivals intruding into each other.

A key time-scale in the temporal evolution is the one associated with the
polaronic tunneling between the two sites. Since this time scale is
intimately connected with the polaron binding energy, we plot in Fig.~6, 
the
logarithm of the tunneling time  as a function of the binding energy. The
solid line denotes the harmonic oscillator limit, $\lambda \rightarrow
\infty $, whereas the dashed line denotes the infinite square-well limit, 
$\lambda =1$. Note that for small values of the binding energy, the
time-scales for both cases is only weakly dependent on the energy. However,
for larger coupling (binding energy), both show a clear linear dependence
(albeit with different slope). This clearly indicates that even for the
box-like potential, the polaronic tunneling time scale is exponentially
dependent on the binding energy. While well-known for harmonic polarons,
this exponential dependence constitutes an important new result for
anharmonic polarons emerging from the present analysis.

\section{Summary}

By analyzing the dynamics and energetics of a quasiparticle interacting 
with
a tunably anharmonic oscillator, specifically described by a %
\mbox{P\"oschl-Teller  } potential, we find that, in the limit of strong
coupling between the quasiparticle and the oscillator, selftrapping is
robust and persists for strong anharmonicities, with the polaron tunneling
time scale being exponentially dependent on the polaron binding energy, a
feature that has been earlier known to be true for harmonic polarons. We
further find that the full quantum result agrees with the predictions of 
the
semiclassical approximation only in this strong-coupling, low-frequency
regime, in agreement with earlier findings for harmonic polarons.

One of us (VMK) acknowledges the financial support of the National Science
Foundation under grant no. DMR-9614848, and of the Los Alamos National
Laboratory under grant no. 0409J0004-3P. 

%
%
%
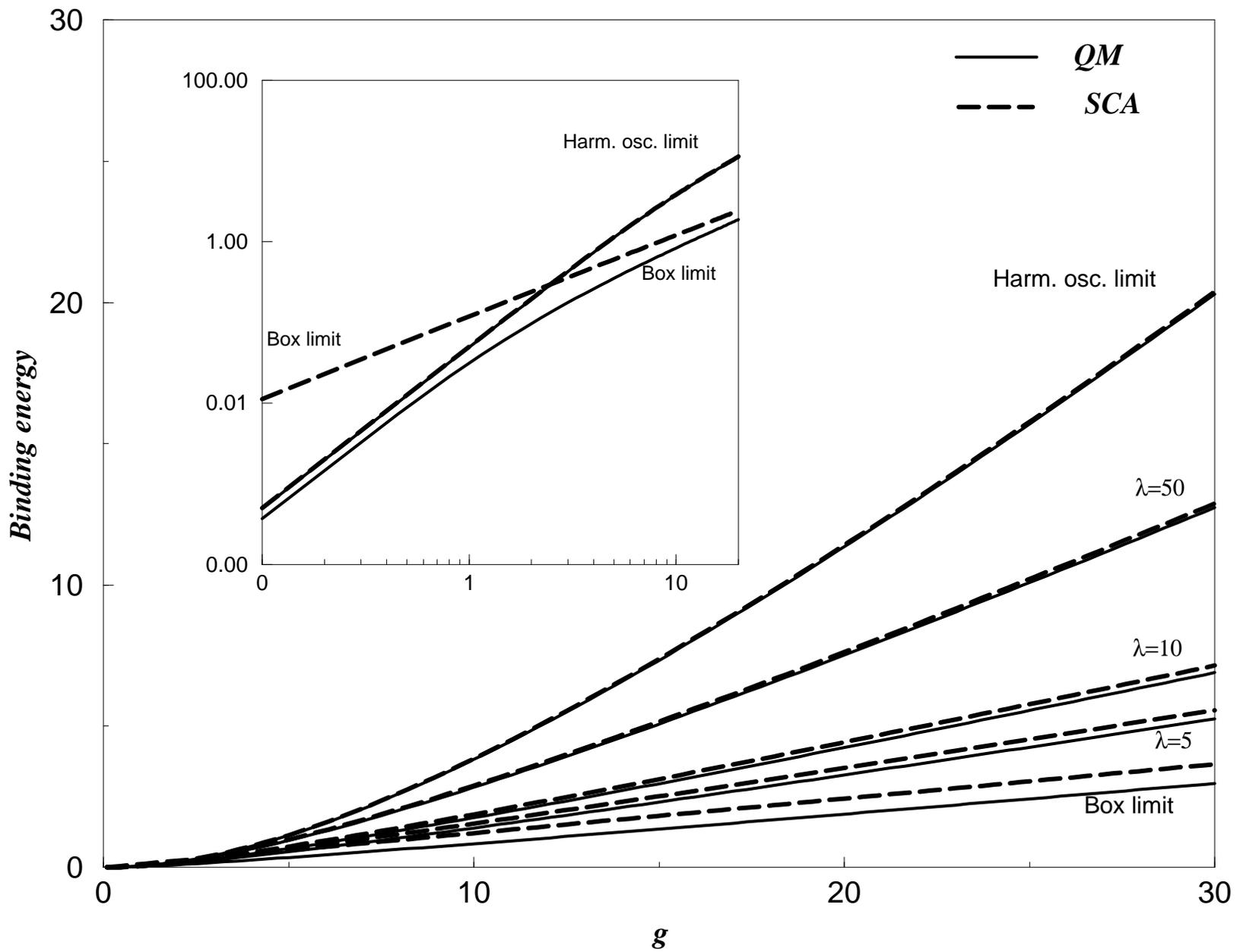
\begin{figure}
\caption{
Polaron binding energy as a function of the quasiparticle-oscillator
coupling constant $g$. The inset shows the same quantities on a logarithmic
scale. The solid line indicates the quantum mechanical result whereas the
dashed line indicates the result of the semiclassical approximation (SCA). 
}
\end{figure}
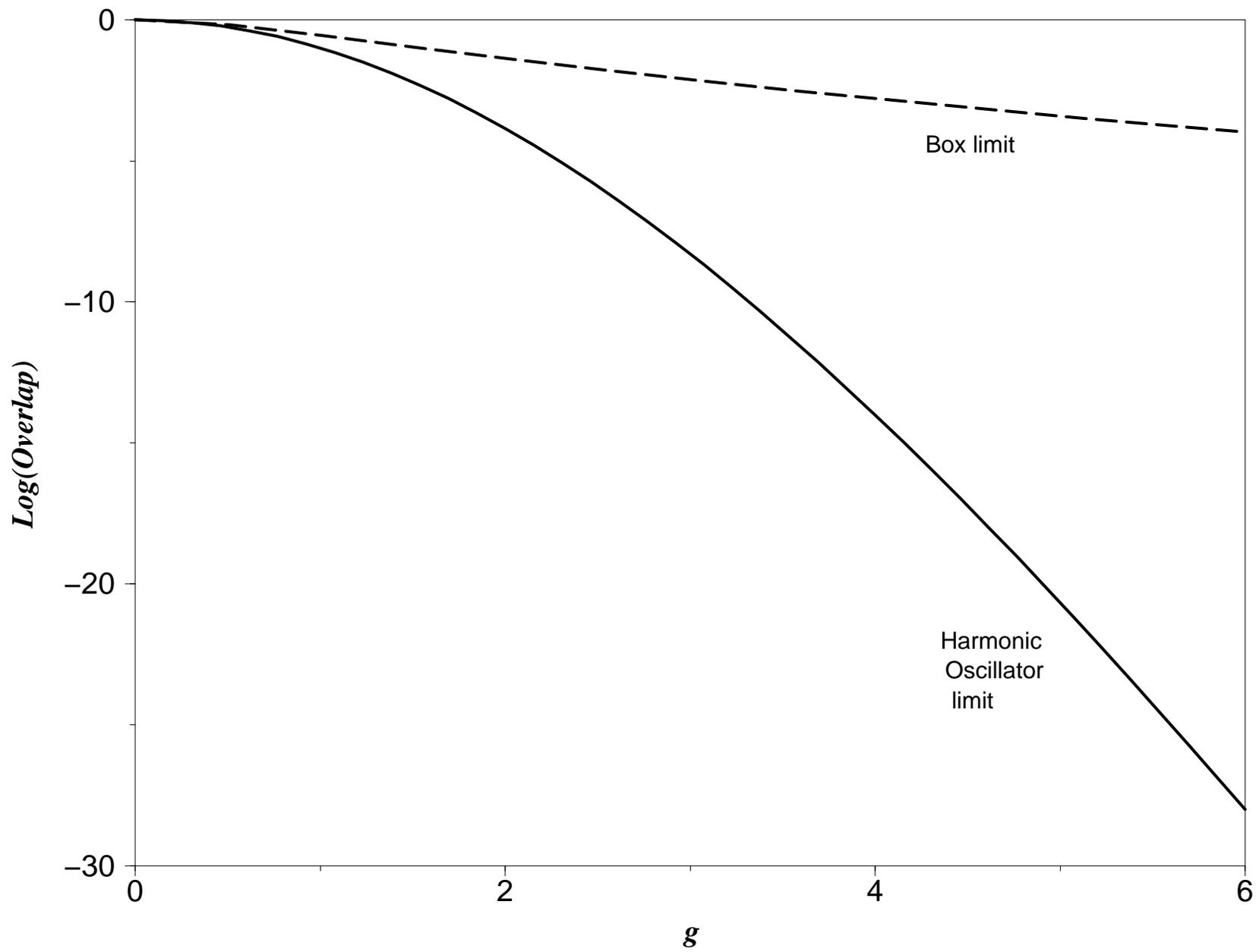
\begin{figure}
\caption{
The overlap of the adiabatically displaced groundstate wavefunctions
plotted logarithmically as a function of $g$ for box limit, $\lambda =1$
(dashed line) and harmonic oscillator limit, $\lambda \rightarrow \infty $
(solid line).}
\end{figure}
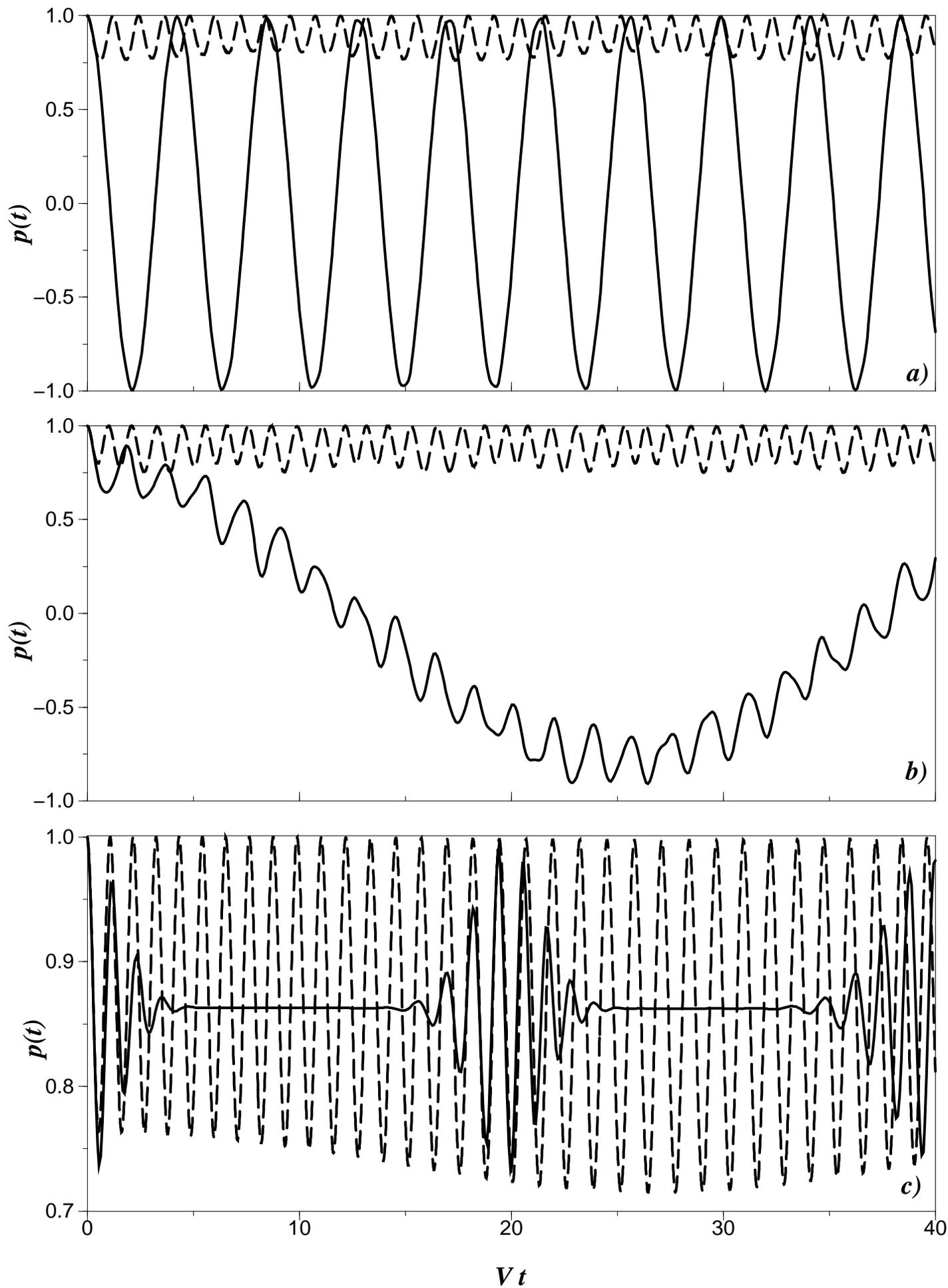
\begin{figure}
\caption{
The evolution of the quasiparticle probability difference $p(t)$ as a
function of dimensionless time $Vt$. In all the figures, the polaron 
binding
energy has been kept fixed at 1.5 V and $\lambda \rightarrow \infty $. The
solid line denotes the fully quantum mechanical result and the dashed line
denotes the result of the SCA. In (a), $\omega _{0}=10V$, (b), $\omega 
_{0}=V
$, (c)= $\omega _{0}=0.1V$.}
\end{figure} 
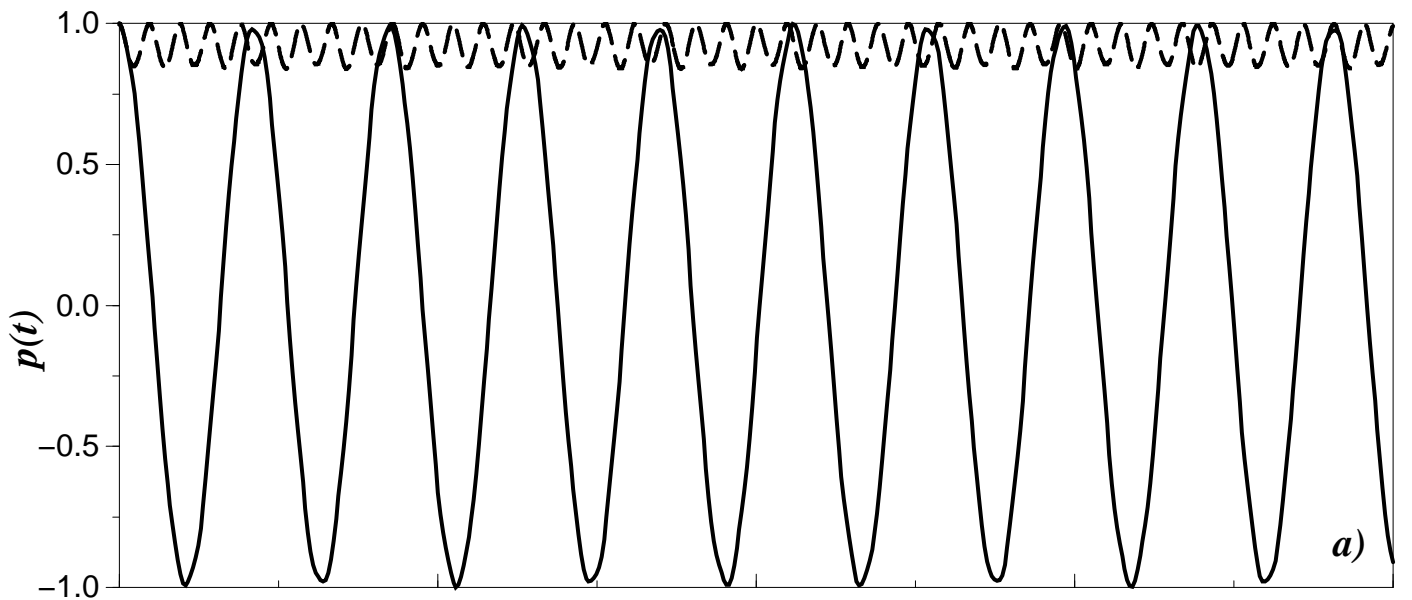
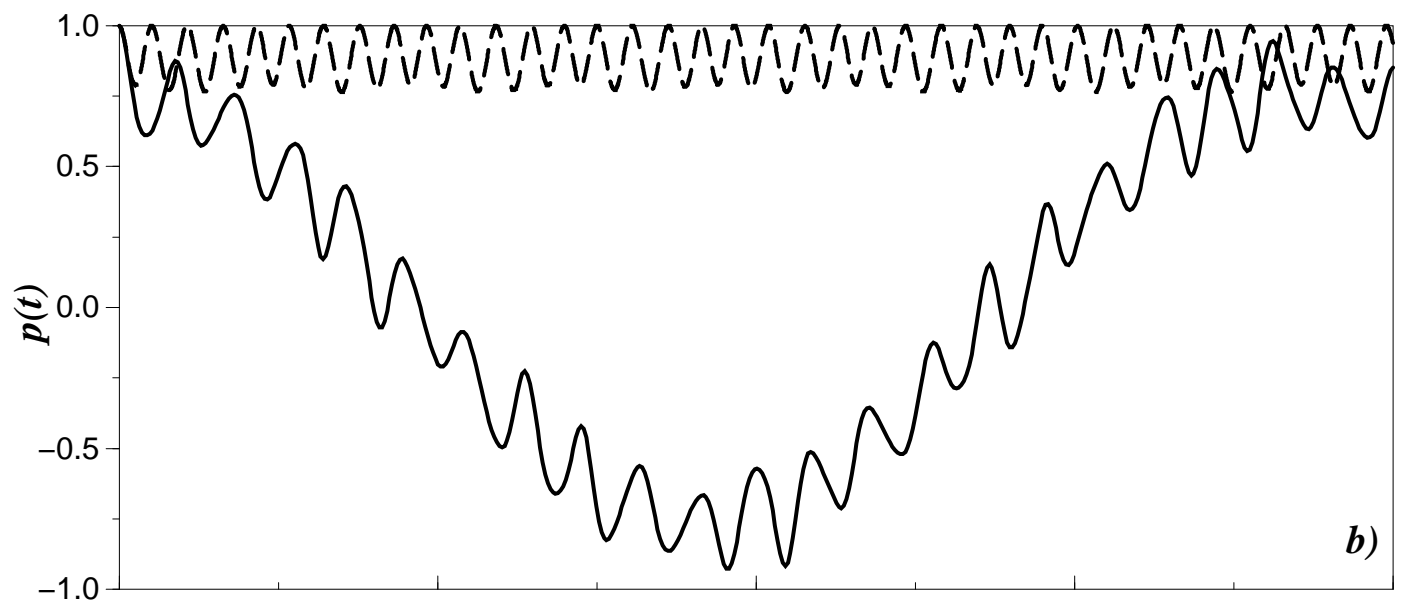
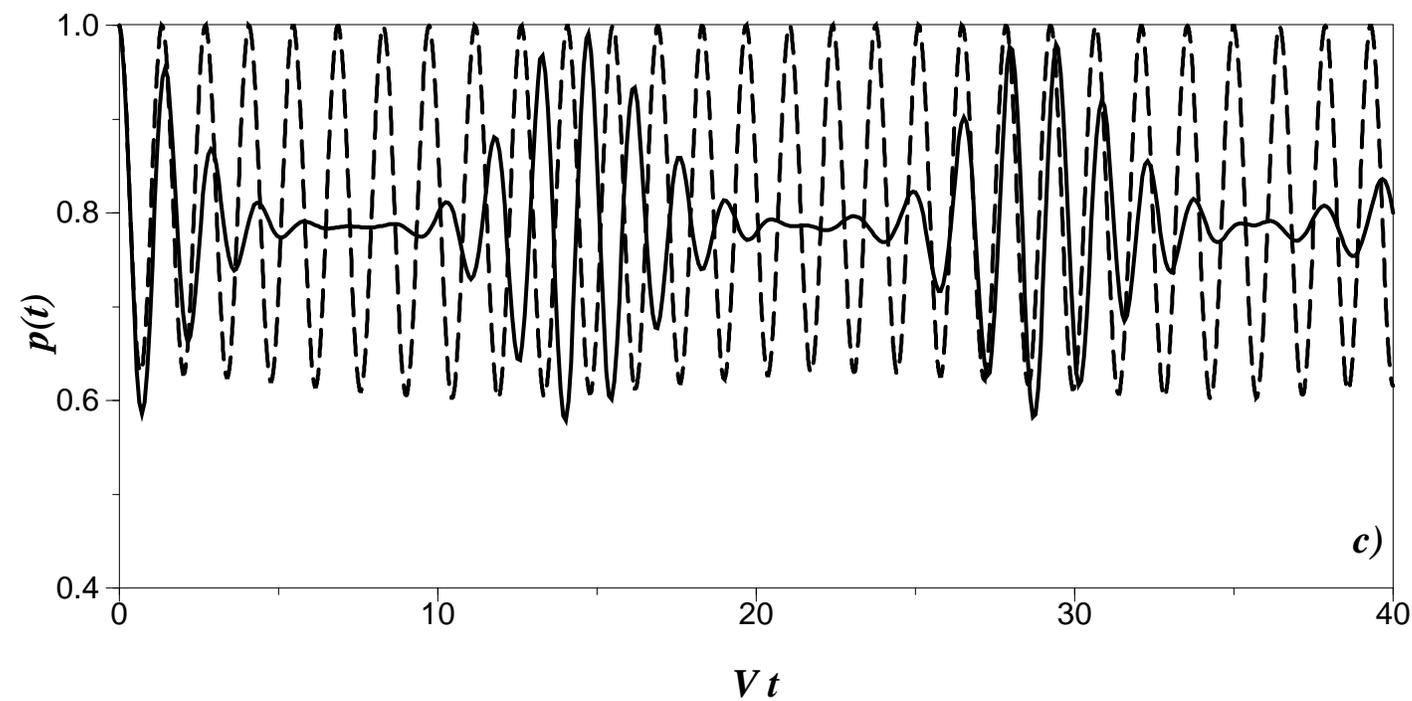
\begin{figure}
\caption{
 Same as Fig.~3, with $\lambda =10$.}
\end{figure} 
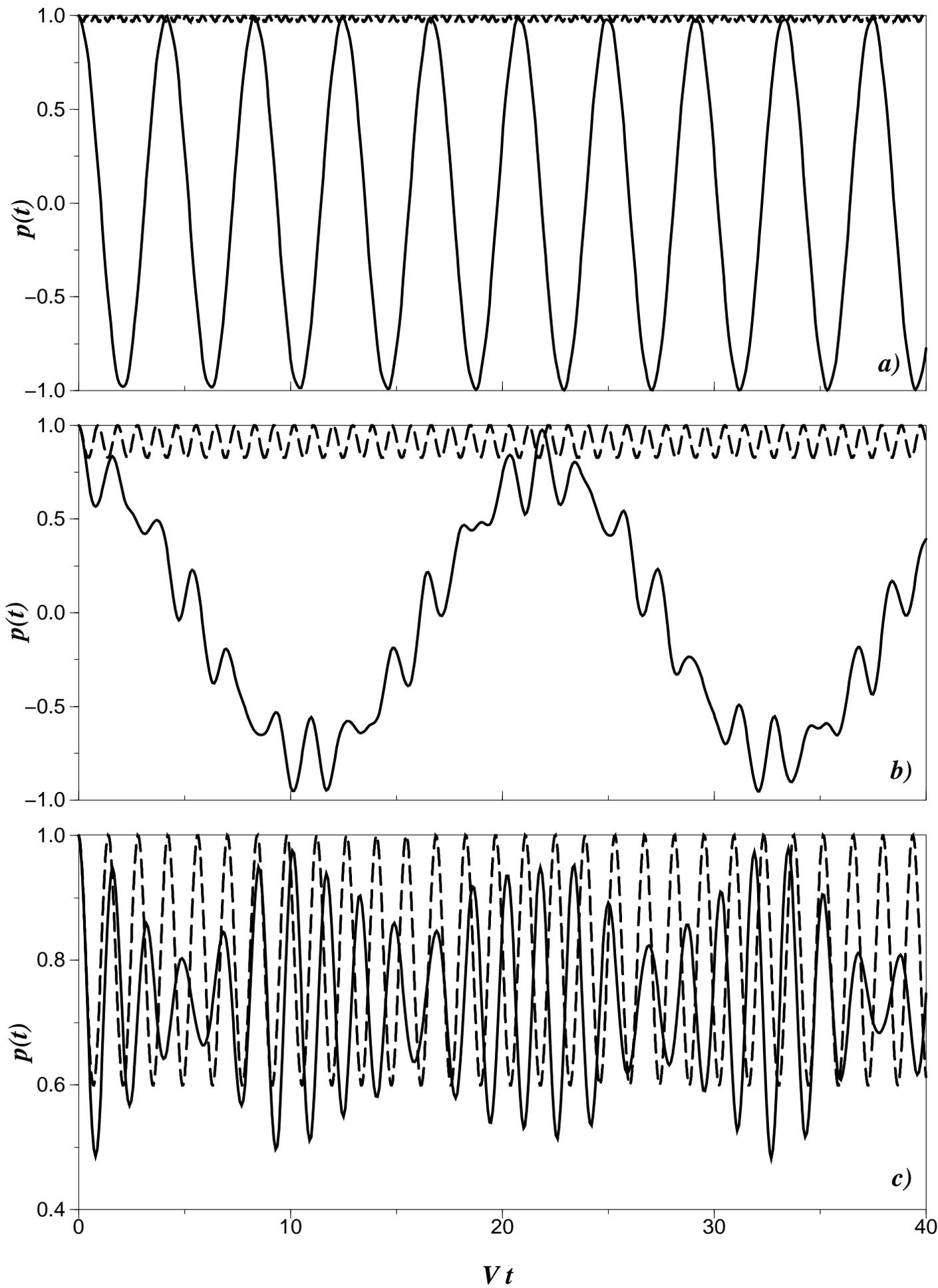
\begin{figure}
\caption{Same as Fig.~3, with $\lambda =1$.}
\end{figure}
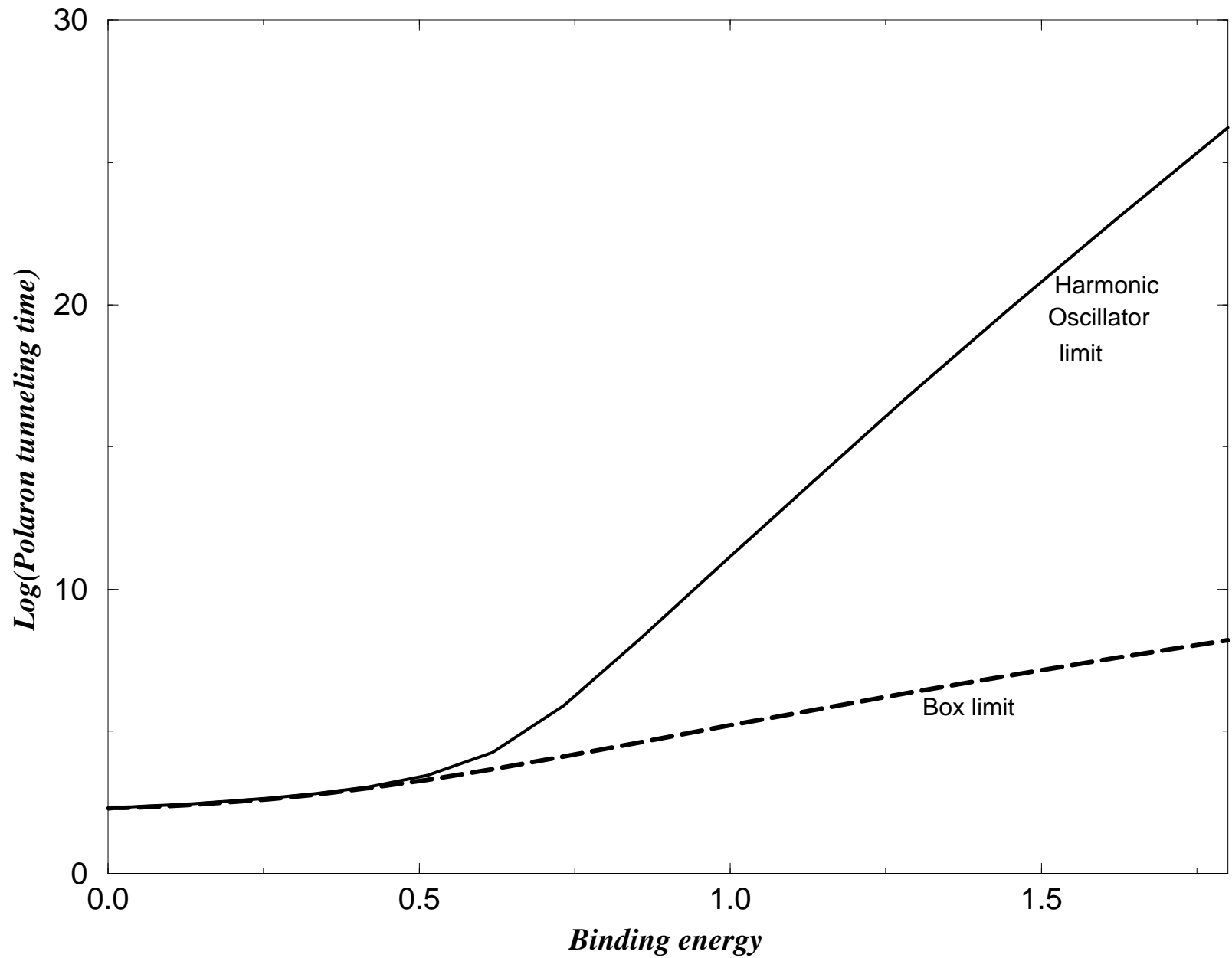
\begin{figure}
\caption{
The polaron tunneling time scale plotted logarithmically as a function 
of
the polaron binding energy, both computed without making the SCA. The solid
line indicates $\lambda \rightarrow \infty $ (harmonic oscillator limit) 
and
the dashed line indicates $\lambda =1$ (box limit).}
\end{figure}
\end{document}